\documentstyle[12pt,psfig]{article}
\begin{document}
\title{Probing the $\rho$ spectral function in hot and dense nuclear matter
by dileptons
\thanks{Work supported by BMBF and GSI Darmstadt.}}
\author{W. Cassing$^1$, E. L. Bratkovskaya$^1$, R. Rapp$^2$ and J. Wambach$^3$
\\[5mm]
$^1$Institut f\"ur Theoretische Physik,
Universit\"at Giessen\\D-35392 Giessen, Germany \\
$^2$ Department of Physics and Astronomy, State University of New York, \\
Stony Brook, NY 11794-3800, U.S.A. \\
$^3$ Insitut f\"ur Kernphysik, TU Darmstadt, Schlo\ss gartenstr. 9,\\
 D-64289 Darmstadt, Germany}
\date{}
\maketitle

\begin{abstract}
We present a dynamical study of $e^+e^-$ and $\mu^+ \mu^-$ production
in proton-nucleus and nucleus-nucleus collisions at CERN-SPS energies on
the basis of the covariant transport approach HSD employing a
momentum-dependent $\rho$-meson spectral function that includes the
pion modifications in the nuclear medium as well as the polarization of
the $\rho$-meson due to resonant $\rho$$-N$ scattering. We find that the
experimental data from the CERES and HELIOS-3 Collaborations can be
described equally well as within the dropping $\rho$-mass scenario.
Whereas corresponding dilepton $q_T$-spectra are found to be very similar,
the inclusive dilepton yield in
the invariant mass range $0.85 \leq M \leq  1.0$~GeV should allow to
disentangle the two scenarios experimentally.
\end{abstract}

\vspace{1cm}


\newpage
\section{Introduction}
The mechanism of chiral symmetry restoration at high baryon density is
of fundamental importance for the understanding of
QCD~\cite{brownrho,chiral}, but a clear experimental evidence has not
been achieved, so far.  Enhanced antikaon yields at SIS energies - as
compared to transport studies using bare kaon
masses~\cite{Fang,Cass96k} - point in this direction, but the present
knowledge on the elementary production cross sections close to
threshold -- especially for the $\Delta N$ production channel -- does
not yet allow for final conclusions. On the other hand, dileptons are
particularly well suited for an investigation of the early phases of
high-energy heavy-ion collisions because they can leave the reaction
volume essentially undistorted by final-state interactions.  Indeed,
dileptons from heavy-ion collisions  have been measured by the DLS
collaboration at the BEVALAC~\cite{ro88,na89,ro89} and by the
CERES~\cite{CERES}, HELIOS~\cite{HELIOS,HELI2}, NA38~\cite{NA38} and
NA50 collaborations~\cite{NA50} at SPS energies.

The recent data on $e^+e^-$ or $\mu^+ \mu^-$ spectra at SPS energies
might provide valuable insight;  Li et al.~\cite{Li} have proposed that
the enhancement of $e^+e^-$ pairs in S~+~Au collisions as observed by
the CERES collaboration~\cite{CERES} stems from an enhanced
$\rho$-meson production (via $\pi^+\pi^-$ annihilation) due to a
dropping $\rho$-mass in the medium. In fact, their analysis -- which is
based on an expanding fireball scenario -- is supported by the
microscopic transport calculations in ref.~\cite{Cass95C}.  Meanwhile
various authors have substantiated the observation of
refs.~\cite{Li,Cass95C}, {\it i.e.}~that the spectral shape of the
dilepton yield is incompatible with 'free' meson form
factors~\cite{Koch96,Srivas,Frankfurt,Soll,Shury}. However, a more
conventional
approach including the change of the $\rho$-meson spectral function in
the medium due to the coupling of $\rho, \pi, \Delta$ and nucleon
dynamics along the lines of refs.~\cite{Herrmann,asakawa,Chanfray,Rapp}
was found to be (roughly) compatible with the CERES data
\cite{Cass95C,Rapp}, as well, whereas the dimuon data of the HELIOS-3
collaboration~\cite{HELIOS} could only be described satisfactorily when
including 'dropping' meson masses~\cite{Cass96H,Li96}. Meanwhile, our
knowledge on the $\rho$ spectral function has improved since -- as
first pointed out by Friman and Pirner~\cite{Friman} -  resonant
$\rho$-$N$ interactions in p-wave scattering significantly enhance the
strength in the vector-isovector channel at low invariant mass.  In
fact, the CERES data for S~+~Au at 200~A~GeV and Pb~+~Au at 160~A~GeV,
using an expanding fireball model, were found to be compatible within
such a hadronic scenario~\cite{RappNPA}.

In the present work we report on novel developments in this direction
by improving on both the reaction dynamics and the $\rho$ spectral
function ({\it e.g.}~by constraints from $\gamma p$ data), which will
be discussed in sect.~2. In sect.~3 the
resulting dilepton spectra are then compared to the data from both CERES and
HELIOS-3 collaborations simultaneously; we also search for observables that
might distinguish between the 'hadronic' and 'dropping mass' scenario,
in particular dilepton $q_T$-spectra and inclusive yields {\em above}
the $\rho$ mass. We finish with a summary and concluding remarks in sect.~4.

\section{Heavy-Ion Reaction Dynamics and in-Medium Rho Propagator}
The dynamical description of proton-nucleus or nucleus-nucleus
collisions is performed within the covariant transport approach
HSD~\cite{Ehehalt} which has been found to account for hadronic data
from SIS to SPS energies reasonably well and has recently been used in
the analysis of dilepton data~\cite{Cass95C,Cass96H,Brat97}.  The
dilepton production is calculated perturbatively by including the
contributions from the Dalitz-decays $\eta \to \gamma l^+l^-$, $\omega
\to \pi^0 l^+l^-$, $\eta^\prime \to \gamma l^+l^-$ and the direct
dilepton decays of the vector mesons $\rho, \omega$ and $\phi$ where
the $\rho$ and the $\phi$ meson may as well be produced in
$\pi^+\pi^-$ and $K \bar{K}$ or $\pi \rho$ collisions, respectively.
For a detailed description of all these processes we refer
the reader to ref.~\cite{Brat97}.  We note that we discard
baryon-baryon ($BB$), meson-baryon ($mB$) and meson-meson ($mm$)
bremsstrahlung channels as well as the Dalitz decays of the baryon
resonances in order to avoid double counting when employing the full
$\rho$ spectral function.

The dilepton radiation from $\rho$ mesons is calculated as
\begin{equation}
\label{rho}
{dN_{l^+l^-}\over dM} = - {2M\over \pi} \ {\rm Im} D_\rho (q_0, q; \rho_B, T)
\end{equation}
where $D_\rho$ is the $\rho$-meson propagator in the hadronic medium,
depending on the baryon density $\rho_B$ and temperature $T$ as well
as on energy $q_0$ and 3-momentum $q\equiv |\vec q|$ in the
local rest frame of the baryon current ('comoving' frame).
The invariant mass $M$ is related to the
$\rho$-meson 4-momentum in the nuclear medium as
\begin{equation}
\label{m}
M^2 = q_0^2 - q^2 \ ,
\end{equation}
while
\begin{equation}
\label{drho}
{\rm Im} D_\rho = \frac{1}{3} ( {\rm Im} D^L_\rho + 2 {\rm Im} D^T_\rho)
\end{equation}
is spin averaged over the longitudinal and transverse part of the
$\rho$-propagator:
\begin{equation}
\label{Imd}
{\rm Im} D^{L,T}_\rho (q_0, q; \rho_B, T) = \frac{{\rm Im}
\Sigma^{L,T}_\rho (q_0, q; \rho_B, T)}{|M^2 - (m^{bare}_\rho)^2 -
\Sigma^{L,T}_\rho (q_0, q; \rho_B, T)|^2} \ .
\end{equation}
The $\rho$-meson selfenergy $\Sigma^{L,T}_\rho (q_0, q; \rho_B, T)$ is
obtained by combining the effects of the different hadronic interactions as:
\begin{equation}
\label{long}
\Sigma^L_\rho = \Sigma_{\rho \pi \pi} + \Sigma^L_{\rho \pi a_1} +
\Sigma^L_{\rho K K_1} + \Sigma^L_{\rho \bar{K} \bar{K}_1}
\end{equation}
\begin{equation}
\label{trans}
\Sigma^T_\rho = \Sigma_{\rho \pi \pi} + \Sigma_{\rho B B^{-1}} +
\Sigma^T_{\rho \pi a_1} + \Sigma^T_{\rho K K_1} +
\Sigma^T_{\rho \bar{K} \bar{K}_1} \ .
\end{equation}
The explicit evaluation of the various selfenergy components is
discussed in detail in ref.~\cite{RappNPA}. We here employ an improved
version in the following respects:
\begin{itemize}
\item[a)] in addition to p-wave
$\rho N \to B$ interactions ($B$=$N$, $\Delta$, $N(1720)$ and $\Delta(1905)$)
we also account for s-wave excitations into $N(1520)$ and $\Delta(1700)$
resonances (as usual, the corresponding coupling constants are estimated
from the $\rho$$N$ partial decay width);
\item[b)] the coupling of the resonances
to virtual photons is calculated within the improved vector dominance
model of Kroll, Lee and Zumino~\cite{KLZ} to avoid overestimates of
the $B\to N\gamma$ branching ratios;
\item[c)] the medium modifications of the
two-pion selfenergy $\Sigma_{\rho\pi\pi}$ are extended to arbitrary
3-momentum within the recently developed model of
Urban {\it et al.}~\cite{Urban}.
\end{itemize}
The combined $\rho$-meson selfenergy is then further constrained by
experimental information on $\gamma p$ and $\gamma A$ absorption cross
sections~\cite{RUBW} which correspond to the limit $M^2\to 0$.
The constraints by photo-production data have been stressed by several
authors \cite{Axel,Brat97,Shury,Li97,Steele1,Steele2} and thus
qualify as stringent benchmarks for the $\rho$ spectral function.
We note that by resummation of hadronic interactions in the $\rho$-meson
selfenergy $\Sigma_\rho$ (eqs.~(\ref{long}),(\ref{trans})) we incorporate
dynamical effects which are not only linear in the baryon density
$\rho_B$, but also of higher order. The resummation of interactions at
fixed baryon density and temperature involves the  assumption that all
interaction terms add coherently -- an assumption that might become
questionable due to the short time scales of a few fm/c in heavy-ion
collisions at SPS energies \cite{Ehehalt}. Our approach thus differs
conceptually from the analysis performed in Refs.~\cite{Steele1,Steele2}
where only dilepton rates linear in $\rho_B$ have been considered and
higher order coherent effects have been discarded.

As an example -- relevant for the heavy-ion reactions at SPS energies --
we show in Fig. 1 the spin averaged ${\rm - Im} D_\rho (q_0, q; \rho_B, T)$
as a function of the invariant mass $M$ and the 3-momentum $q$ for a
temperature of 150 MeV at $\rho_B = 0, \rho_0, 2 \rho_0$ and $3
\rho_0$, respectively. With increasing baryon density we find the
$\rho$ spectral function to increase substantially in width showing
only minor structures at high density. Thus the lifetime of the
$\rho$-meson in the nuclear medium becomes very short due to well
established hadronic interactions. As a consequence, its average
propagation in space is limited to distances less than 0.5~fm already
at normal nuclear matter density (even for high momenta).

The cross section for the pion annihilation channel
$\pi^+ \pi^- \rightarrow \rho^0 \rightarrow e^+ e^-$ is
taken in line with ref.~\cite{PKoch93} as
\begin{eqnarray}
\sigma_{\pi^+\pi^-\to e^+e^-}(M) = - {16\pi^2\alpha^2\over g_{\rho\pi\pi}^2} \
 {1\over k^2 M^2} \ (m_\rho^{bare})^4 \ {\rm Im} D_\rho (q_0, q; \rho_B, T)
\ , 
\label{pipian}\end{eqnarray}
where $k=(M^2-4m_\pi^2)^{1/2}/2$ is the pion 3-momentum in the
center-of-mass frame and $\alpha$ is the fine structure constant. The
$\rho\pi\pi$ coupling constant and bare $\rho$ mass are fixed to
reproduce the pion electromagnetic form factor in free space~\cite{RappNPA}.

In this way we ensure that the Fourier transformation of the isovector
current-current correlation function in vacuum is consistent with the
experimental information in terms of the vacuum $\rho$ spectral
function \cite{Steele1}. We note that constraints from chiral
symmetry are not fully accounted for in our model for the spectral function 
since the $a_1(1260)$ meson is not treated as the chiral partner of the 
$\rho$ meson~\cite{Shury}. However, the dominant effects arise from  
p-wave $\pi N$- and resonant $\rho N$
scattering, where chiral symmetry does not play a dominant role.

Our dynamical calculations are carried out as follows: the time-evolution
of the proton-nucleus or nucleus-nucleus collision is described within the
covariant transport approach HSD~\cite{Ehehalt,Brat97} without dropping
vector meson masses. Whenever a $\rho$-meson is produced in the course of
the hadronic cascade (by baryon-baryon, meson-baryon or pion-pion collisions),
its 4-momentum in the local rest frame of the baryon current
is recorded together with the information on the local baryon density,
the local 'transverse' temperature  and  its production source. We note that
the definition of a local temperature is model dependent; here we have used
a logarithmic fit to the transverse $p_t$ spectra of mesons at midrapidity.
Without going into a detailed discussion of this issue we note that our
results
for dilepton spectra do not change within the numerical accuracy when using
a constant temperature $T = 150$~MeV at SPS energies since
${\rm Im} D_\rho$ depends rather weakly on $T$ (at a given nucleon density).

\section{Comparison to HELIOS-3 and CERES Data}
We begin our discussion of the
dilepton spectra at SPS energies with the p~+~W system at 200~GeV incident
energy. Fig.~2 shows the
spectral channel decomposition as a function of the $\mu^+\mu^-$
invariant mass $M$ in comparison to the data of
the HELIOS-3 collaboration~\cite{HELIOS} including the experimental
cuts in transverse mass and rapidity as well as the experimental mass
resolution as in ref.~\cite{Cass96H}. The spectrum is normalized to the
charged particle multiplicity $N_c$ in the pseudorapidity bin 3.7 $\leq
\eta \leq $ 5.2.  The upper part of Fig.~2 shows the dimuon spectra
when employing the 'free' $\rho$ spectral function whereas the
lower part displays our results when accounting for the full medium
modifications of the $\rho$-meson propagator. As is apparent from
Fig.~2, the spectrum for p~+~W can be fully accounted for by the
electromagnetic decays of the $\eta, \eta^\prime$ and vector
mesons $ \rho^0, \omega$ and $\phi$; contributions from meson-meson
channels ($\pi^+\pi^-, K\bar K, \pi \rho$) are of minor importance here.
Furthermore, since the $\rho$-meson essentially hadronizes in the
vacuum, there is no noticeable difference between the summed spectra
in the upper and lower part of Fig.~2.

The situation changes appreciably when turning to nucleus-nucleus
collisions.  In Fig.~3 we compare the results of our calculation for
the differential dilepton spectra for S~+~W at 200~A$\cdot$GeV with the
experimental data~\cite{HELIOS} employing the same cuts and mass
resolution as before. Again, the upper part shows our results for the
'free' $\rho$ spectral function whereas the lower part represents our
results including the full medium modifications of the $\rho$
propagator. Even at forward rapidities as measured in central S~+~W
(200~A$\cdot$GeV) a sizable contribution for invariant masses 0.3~GeV
$\leq M$ $\leq$ 0.7~GeV stems from the $\pi^+\pi^- \rightarrow \rho
\rightarrow \mu^+ \mu^-$ channel. Also in the $\phi$ mass regime around
1~GeV there is a significant contribution from $K\bar K$ and $\pi \rho$
annihilation to dimuons which explains the enhanced $\phi/(\rho +
\omega)$ ratio in S~+~W relative to p~+~W. As has been found
earlier~\cite{Cass96H,Li96} the experimental spectrum cannot be
properly described in terms of a 'free' $\rho$ spectral function: there
is an excess for invariant masses $M\approx$~0.8~GeV and insufficient
yield around $M\approx$~0.5~GeV. The description of the data is
significantly improved when including the hadronic medium modifications
of the $\rho$ propagator (lower part of Fig.~3) -- quite analogous to
the 'dropping mass' scenario~\cite{Cass96H} when employing a
$\rho$-mass shift according to Hatsuda and Lee~\cite{hatsuda}.
However, there persists a significant difference: whereas in the
'dropping mass' scenario (cf.~Fig.~4 of ref.~\cite{Cass96H}) the dimuon
yield is underestimated for $M \geq$ 1.2 GeV the in-medium $\rho$
spectral function seems to give sufficient pairs in this region due to
its substantial broadening in dense matter (cf. Fig.~1).

In Fig.~4 we also compare the results of our calculation for the
differential dilepton spectra in Pb~+~Au at 160~A$\cdot$GeV and
$b=5$~fm with preliminary experimental data~\cite{Ullrich,Drees96}. The
$e^+e^-$ acceptance cuts in pseudorapidity ($2.1 \le \eta \le 2.65$),
transverse momentum ($p_T \ge 0.175$~GeV/c) as well as in the opening
angle of the $e^+e^-$ pair ($\Theta \ge 35$~mrad) are taken into
account.  Furthermore, the experimental mass resolution has been
included in evaluating the final mass spectrum, which is normalized to
the number of charged particles $dn_{ch}/d\eta$ in the pseudorapidity
bin 2.1 $\leq \eta \leq$ 3.1.  Again, the upper part shows our results
for the 'free' $\rho$ spectral function whereas the lower part
represents our results including the full medium modifications.  For
Pb~+~Au at 160~A$\cdot$GeV (and semi central collisions) the dominant
yield for invariant masses $0.3\leq M \leq$ 0.7~GeV stems from
$\pi^+\pi^-$ annihilation.  In the $\phi$ mass regime around 1~GeV
there is again a large contribution from $K\bar K$ and $\pi \rho$
annihilation to dileptons.  Within the present statistics, however,
there is no unique signal for in-medium effects since the calculation
with the 'free' spectral function (upper part) also describes the data
except for one point at 0.7~GeV (cf.~also ref.~\cite{Frankfurt}). The
calculation with the full spectral function describes the data somewhat
better to the same accuracy as within  the 'dropping mass'
scenario~\cite{Brat97}. We note that the comparison with the data in
Fig.~4 has been performed for impact parameter $b = 5$~fm, for which
our transport model gives a charged particle multiplicity
$dn_{ch}/d\eta \approx$ 260 in accordance with the experimental
normalization.

Since the present dilepton invariant mass spectra do not discriminate the
different models we also explore if the momentum distributions of the
lepton pairs give some further insight (as suggested by Friman and
Pirner~\cite{Friman} and in ref.~\cite{Brat97}). Our calculated
transverse momentum spectra -- integrated over rapidity and the
invariant mass range $0.3 \le M \le 0.7$~GeV --  are shown in Fig.~5 for
Pb~+~Au at 160~A$\cdot$GeV and $b=2$~fm using $q_T$-bins of
50~MeV/c.  The dashed line in Fig.~5 displays the result with the
'free' spectral function while the solid line shows the result for the
in-medium spectral function. Apart from an overall increase of the
spectrum we do {\em not} observe significant changes of the slope
in $q_T$ -- as was the case for the 'dropping mass' scenario
(cf.~Fig.~13 of ref.~\cite{Brat97}). Thus, the $q_T$ spectra do not
 qualify in disentangling the different schemes either.

Another signature studied in ref.~\cite{Brat97} and also suggested
in \cite{Steele2} is related to the total
dilepton yield between the free $\omega$ and $\phi$ peaks where the
dominant contribution stems from $\rho$ decays. In this regime the
'dropping mass' scenario leads to a {\em reduction} by about a factor
2-3 as compared to a 'free' $\rho$ spectral function for Pb~+~Pb at
160~A$\cdot$GeV~\cite{Brat97}, while a broadening of the $\rho$
spectral function due to hadronic interactions shows an {\em
enhancement}. This feature is demonstrated quantitatively in Fig.~6 for
Pb~+~Au at $b$=5~fm for the different scenarios employing an
experimental mass resolution of $\Delta M$=10~MeV.
We find a difference by about a factor of 5 between the 'dropping mass'
scenario (dashed line) and the hadronic spectral function approach
(solid line) which might be accessible experimentally with the CERES
detector at improved mass resolution. We note, furthermore, that in the
spectral function approach the relative enhancement is most pronounced
at $M\simeq 0.3$~GeV while in the dropping mass scheme we find a maximum
enhancement at $M\simeq 0.5$~GeV for this system.

\section{Summary}
On the basis of the covariant
transport approach HSD~\cite{Ehehalt,Brat97} we have studied dilepton
production in proton and heavy-ion induced reactions at SPS energies.
Various contributions are accounted for:
the Dalitz-decays of  $\eta, \omega, \eta^\prime$ mesons,
the direct dilepton decays of the vector mesons $\rho, \omega$
and $\phi$ as well as $K\bar K$ and $\pi \rho$ channels. It is found
that for p~+~W collisions at 200~GeV the mesonic decays almost
completely determine the dilepton yield, whereas in S~+~W and Pb~+~Au
reactions the $\pi^+\pi^- \rightarrow \rho \rightarrow l^+ l^-$, $K\bar K
\rightarrow \phi \rightarrow l^+ l^-$ annihilation channels and
$\pi \rho$ collisions contribute substantially.  The experimental data
taken by the HELIOS-3 \cite{HELIOS} and CERES
collaborations~\cite{CERES,Ullrich,Drees96} are in general underestimated
by the
calculations for invariant masses 0.35~GeV~$\leq M \leq 0.65$~GeV when
using a 'free' $\rho$ spectral function (in accordance with
refs.~\cite{Li,Koch96,Li96}).

It turns out that both the 'dropping mass' scenario -- using either
Brown/Rho scaling \cite{Li,Li96} or dropping vector meson masses
according to QCD sum rule studies \cite{Cass95C,Cass96H,Brat97} -- as
well as the hadronic spectral function approach~\cite{RappNPA} lead to
dilepton spectra that are in good agreement with the experimental data
for all systems at SPS energies (cf. also ref.~\cite{Drees96}).  In
order to discriminate between the different models we have
investigated dilepton transverse momentum distributions for invariant
masses $0.3 \leq M \leq 0.7$~GeV in Pb~+~Au collisions; unfortunately,
only little differences have been found as compared to the dropping
mass scenario~\cite{Brat97}.  However, the in-medium broadening of the
$\rho$ spectral function results in a much higher dilepton rate than
the dropping mass scenario {\em above} the $\omega$-mass. Therefore one
might experimentally distinguish the two different scenarios by the
total dilepton yield in the invariant mass range $0.8 \leq M \leq
1$~GeV in Pb~+~Au collisions (cf.~Fig.~6), provided that the
experimental mass resolution is on the order of about 10~MeV.

\vspace{1cm}

\noindent
{\bf Acknowledgement}

The authors are grateful for many helpful discussions with
G.E.~Brown, A.~Drees, C.~M.~Ko and  U.~Mosel. We are indebted to M.~Urban
for providing us with results on the finite 3-momentum dependence in the
two-pion selfenergy. One of us (RR) acknowledges support from the
Alexander-von-Humboldt foundation as a Feodor-Lynen fellow. This work was
supported in part by the U.S. department of energy under contract No.
DE-FG02-88ER40388 and by the National Science Foundation, NSF PHY 94-21309 .

\newpage
\noindent{\large\bf Figure Captions}\\[2mm]

\begin{itemize}  

\item[Fig.~1:] The (negative) imaginary part of the $\rho$ propagator averaged
over the
longitudinal and transverse components as a function of the invariant mass
$M$ and the
momentum $q$ for baryon densities of 0, 1 $\rho_0$, 2 $\rho_0$, and 3
$\rho_0$ and temperature $T = 150$~MeV. Note the different absolute
scales in the individual figures.

\item[Fig.~2:] The calculated dimuon spectra (full solid line) for p~+~W at
200~A$\cdot$GeV in comparison with the data from
ref.~\protect\cite{HELIOS}. The thin lines indicate the individual
contributions from the different production channels including the
HELIOS-3 acceptance and mass resolution; {\it i.e.}~ starting from low
$M$: $\eta \to \gamma e^+ e^-$ (dashed line), $\omega \to \pi^0 e^+ e^-$
(dot-dashed line), $\eta^\prime \to \gamma e^+ e^-$ (long dashed line);
for $M \approx $ 0.8 GeV:  $\omega \to e^+e^-$ (dot-dot-dashed line),
$\rho^0 \to e^+e^-$ (dashed line), $\pi^+ \pi^- \to \rho \to e^+e^-$
(dot-dashed line); for $M \approx $ 1 GeV: $\phi \to e^+e^-$
(dot-dashed line), $\pi \rho \to e^+e^-$ (dot-long dashed line), $K
\bar{K} \to e^+e^-$ (dashed line).  The upper part shows the result for
a 'free' $\rho$ spectral function whereas the lower part is obtained
for the full $\rho$ spectral function.

\item[Fig.~3:] Dimuon invariant mass spectra for central collisions of S~+~W
at 200~A$\cdot$GeV (full solid lines) in comparison to the data of the
HELIOS-3 collaboration~\protect\cite{HELIOS}. The upper part shows the
results of a calculation with a 'free' $\rho$ spectral function whereas
the lower part includes the full $\rho$ spectral function as described
in the text.  The assignment of the individual contributions is the
same as in Fig.~2.

\item[Fig.~4:] Dielectron invariant mass spectra for semi central 
collisions of
Pb~+~Au at 160~A$\cdot$GeV (full solid lines) in comparison to the
preliminary data of the CERES collaboration~\protect\cite{Ullrich}. The
upper part shows the results of a calculation with a 'free' $\rho$
spectral function whereas the lower part includes the full $\rho$
spectral function as described in the text.  The assignment of the
individual contributions is the same as in Fig.~2.

\item[Fig.~5:] The dielectron transverse momentum distribution for the
invariant mass range $0.3 \le M \le 0.7$~GeV for Pb~+~Au at 160~A$\cdot$GeV
at $b=5$~fm for the 'free' $\rho$ spectral function (dotted line) and
with our in-medium spectral function (solid line).

\item[Fig.~6:] The dielectron invariant mass spectra for Pb~+~Au at
160~A$\cdot$GeV at $b=5$~fm for the 'free' $\rho$ spectral function
(dotted line) and with our in-medium spectral function (solid line).
The dashed line corresponds to the 'dropping mass' scenario
investigated in ref.~\protect\cite{Brat97}.

\end{itemize} 


\newpage

\psfig{figure=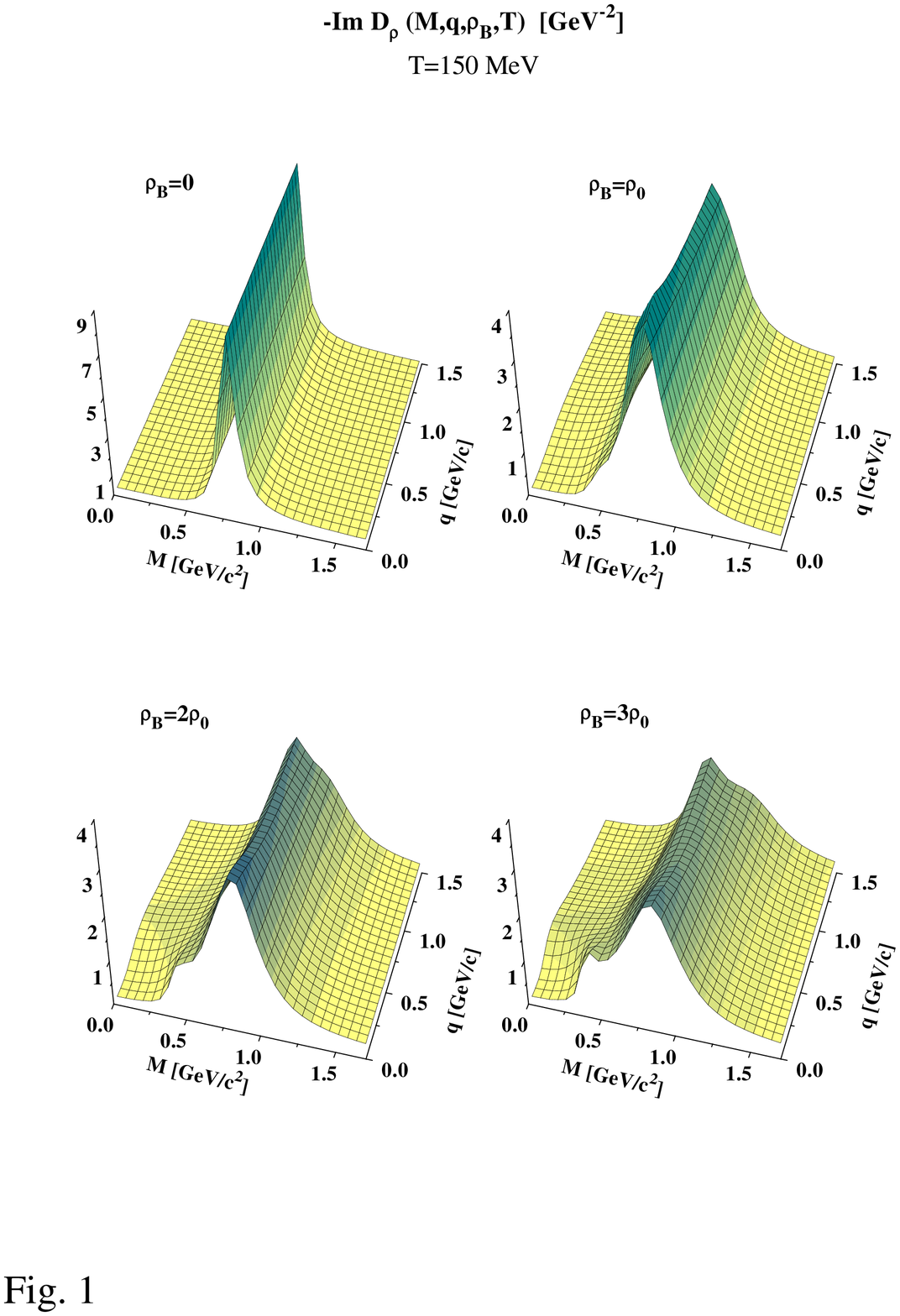,width=16cm,height=22cm}
\psfig{figure=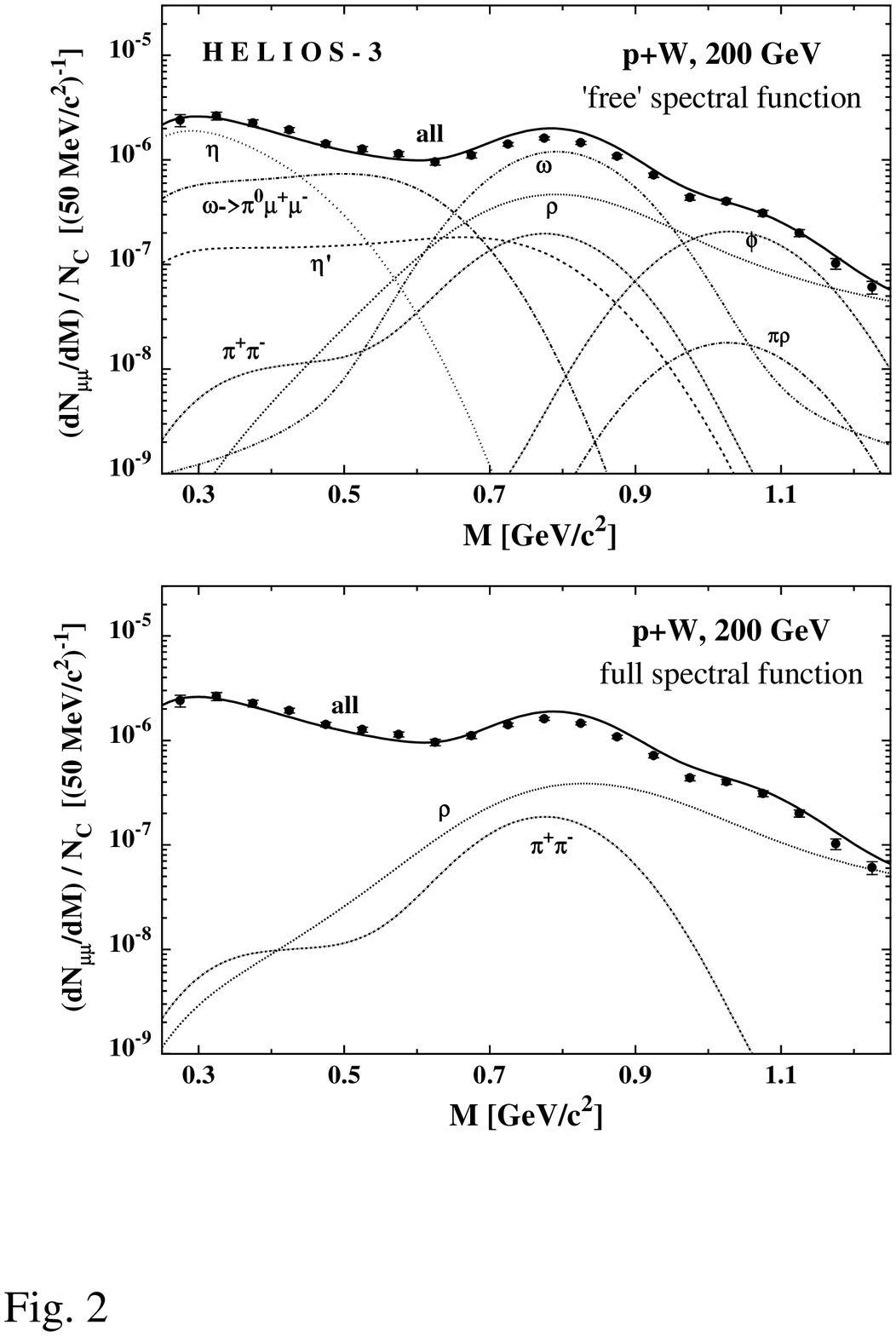,width=16cm,height=22cm}
\psfig{figure=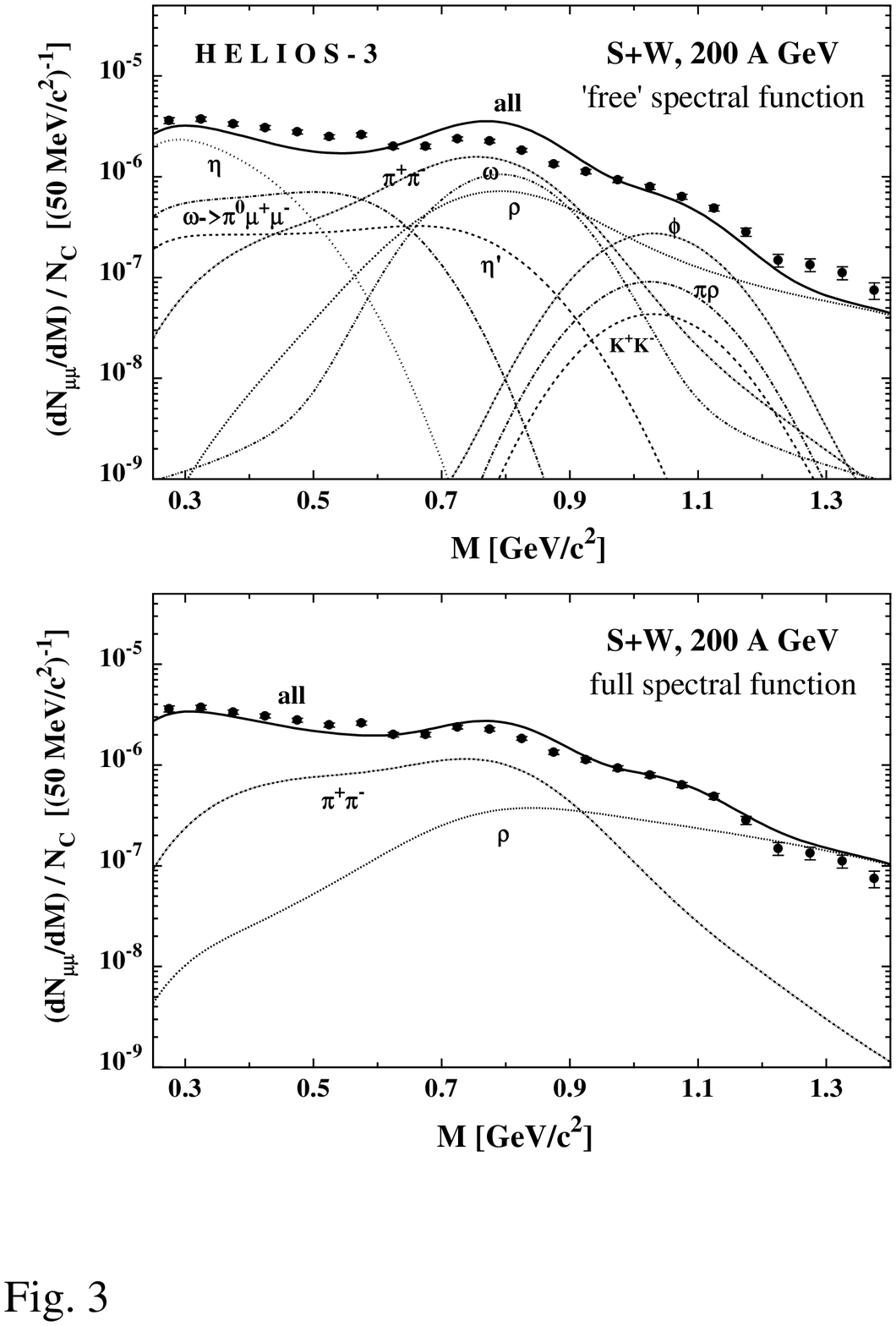,width=16cm,height=22cm}
\psfig{figure=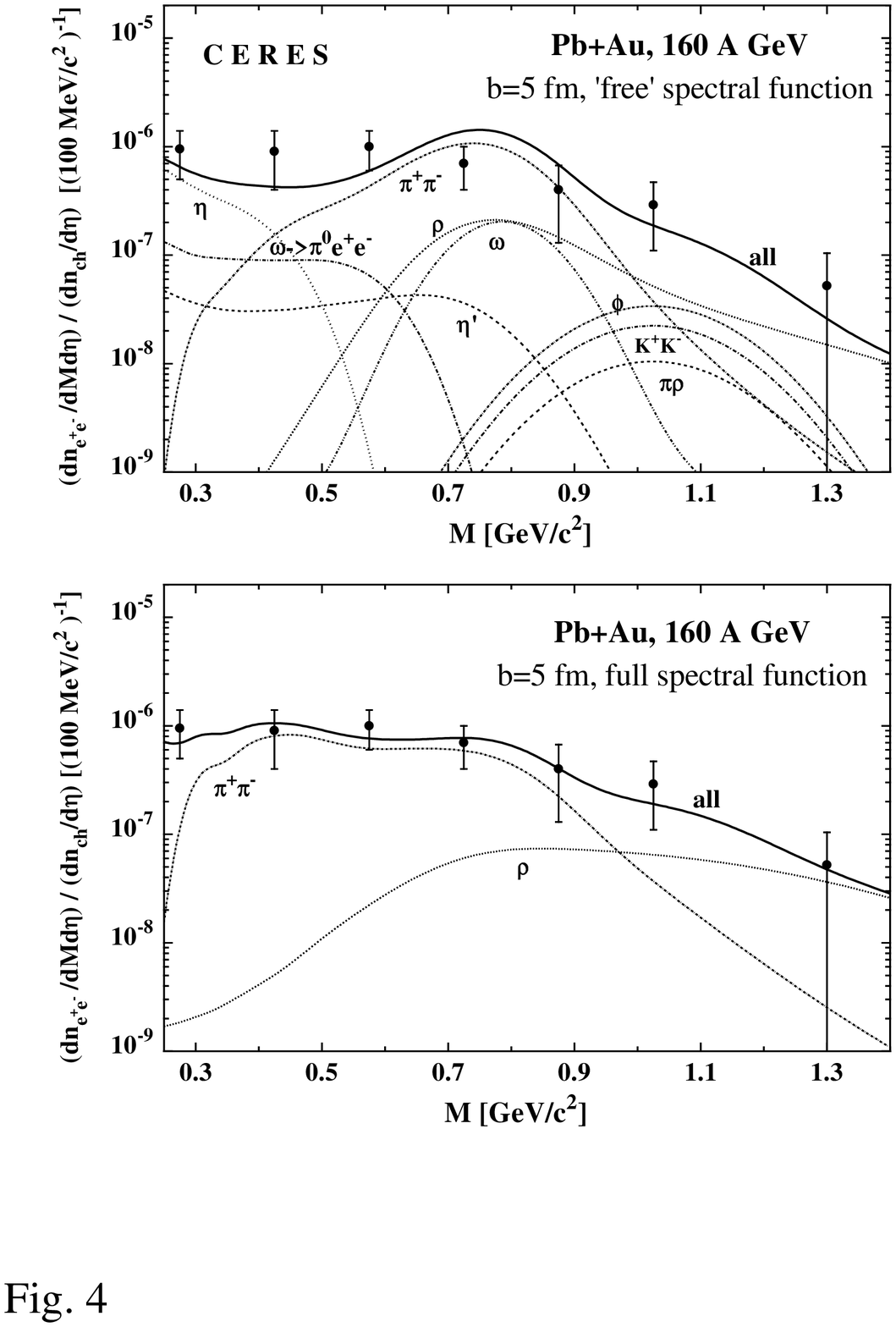,width=16cm,height=22cm}
\psfig{figure=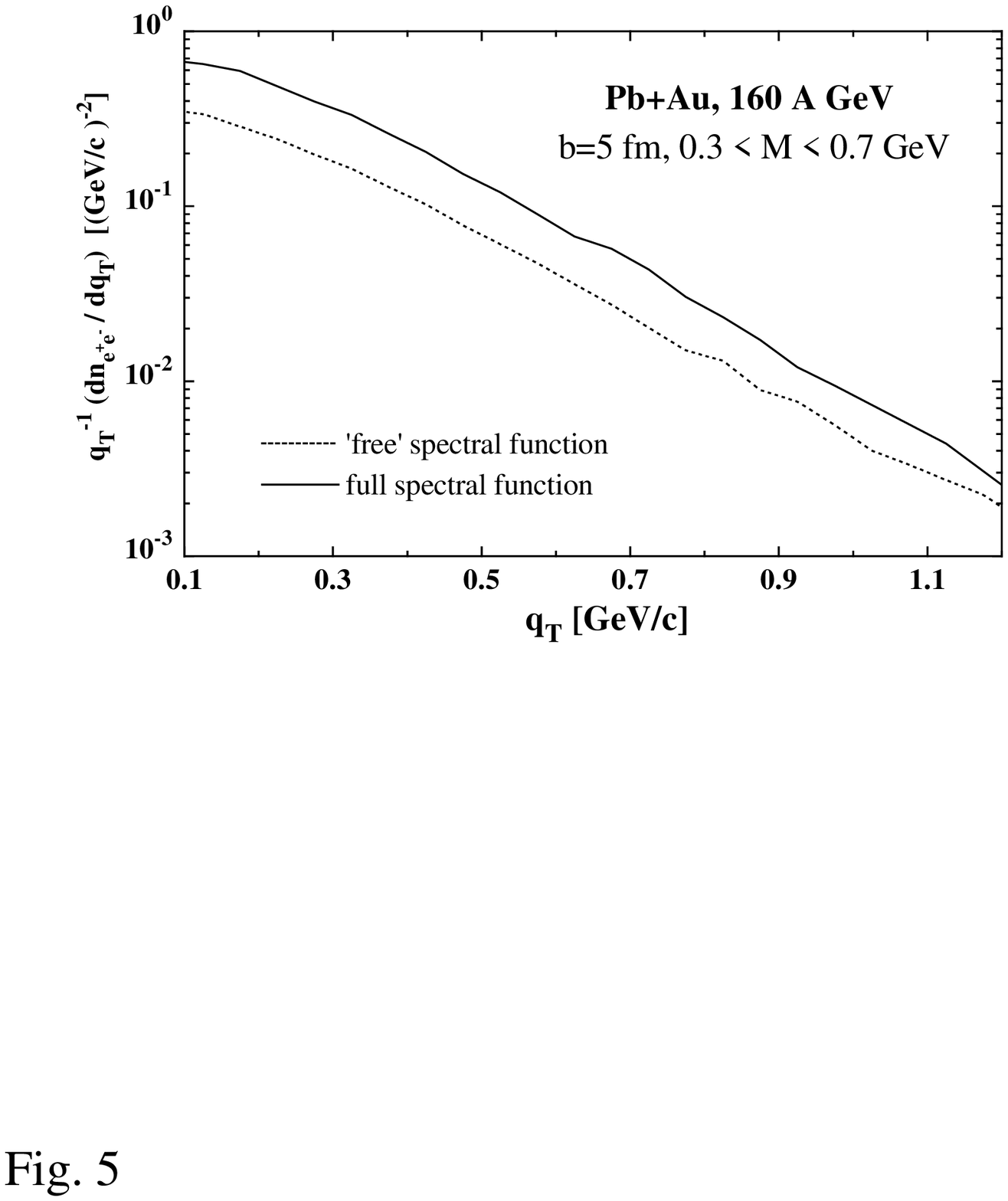,width=16cm,height=22cm}
\psfig{figure=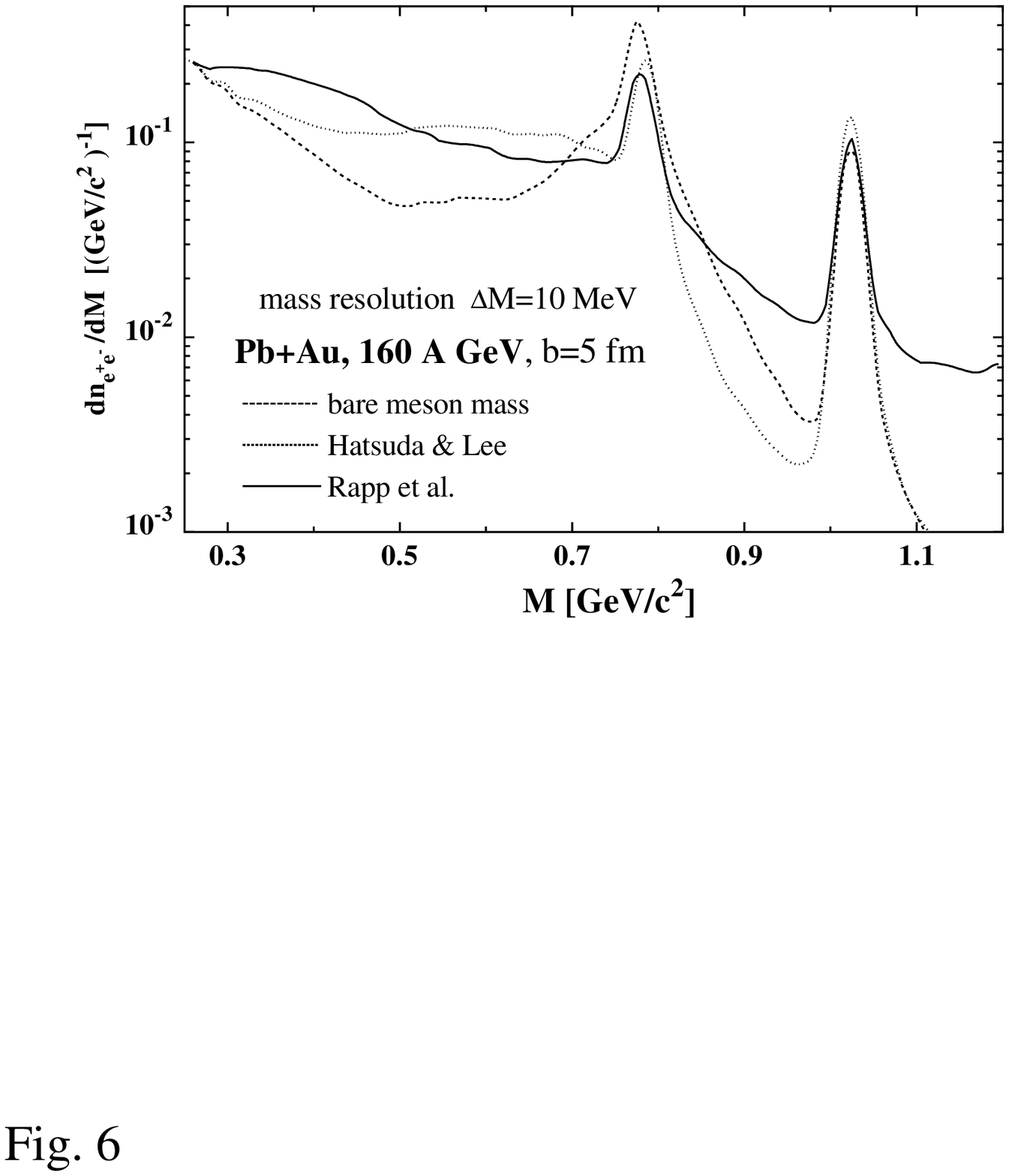,width=16cm,height=22cm}

\end{document}